\documentclass[twocolumn,aps,showpacs]{revtex4}
\usepackage{amsmath}
\usepackage{amssymb}
\usepackage{graphicx}
\usepackage{dcolumn}

\begin{document}

\title{Magnetophonon oscillations of thermoelectric power and combined resonance
in two-subband electron systems}

\author{A. D. Levin,$^1$ G. M. Gusev,$^1$ O. E. Raichev,$^2$
and A. K. Bakarov$^{3,4}$}

\affiliation{$^1$Instituto de F\'{\i}sica da Universidade de S\~ao
Paulo, 135960-170, S\~ao Paulo, SP, Brazil}

\affiliation{$^2$Institute of Semiconductor Physics, NAS of
Ukraine, Prospekt Nauki 41, 03028 Kyiv, Ukraine}

\affiliation{$^3$Institute of Semiconductor Physics, Novosibirsk
630090, Russia}

\affiliation{$^4$Novosibirsk State University, Novosibirsk 630090,
Russia}

\date{\today}

\begin{abstract}
By measuring the thermoelectric effect in high-mobility quantum wells with
two occupied subbands in perpendicular magnetic field, we detect magnetophonon
oscillations due to interaction of electrons with acoustic phonons. These
oscillations contain specific features identified as combined resonances
caused by intersubband phonon-assisted transitions of electrons in the
presence of Landau quantization. The quantum theory of phonon-drag
magnetothermoelectric effect, generalized to the case of multi-subband
occupation, describes our experimental findings.
\end{abstract}


\maketitle

\section{Introduction}

It has been recently established that magnetotransport coefficients of two-dimensional
(2D) high-mobility electron gas in quantum wells (QWs) demonstrate magnetophonon oscillations
(MPO) due to interaction of electrons with acoustic phonons [1-16]. These oscillations
are caused by a combined effect of Landau quantization in the perpendicular magnetic
field $B$ and sensitivity of electron-phonon scattering probability to wavenumbers of
acoustic phonon modes dictated by the kinematics of scattering near the Fermi surface.
The backscattering processes, when the phonon wavenumber $Q$ is close to the Fermi
circle diameter $2k_F$, have a maximum probability. On the other hand, the Landau
quantization implies that the highest scattering probability is realized when phonon
frequency is a multiple of the cyclotron frequency $\omega_c=|e|B/mc$. Since the
acoustic phonon frequency is given by a linear relation $\omega_{\lambda {\bf Q}}=
s_{\lambda} Q$, where $s_{\lambda}$ is the sound velocity of the mode $\lambda$, the
transport is enhanced under the {\em magnetophonon resonance} conditions $2 k_F s_{\lambda}=
n \omega_c$, where $n$ is an integer. As the magnetic field changes, different Landau
levels enter the resonance, and the $1/B$-periodic oscillating picture appears. These
oscillations are not sensitive to the position of the Fermi level with respect to Landau
levels, so they are much more robust to increasing temperature $T$ than the Shubnikov-de
Haas oscillations. Moreover, the amplitude of the oscillations increases with $T$ in
the Bloch-Gruneisen region $T < 2 p_F s_{\lambda}$ due to increase in the number of
phonons contributing to the electron-phonon collisions.

The acoustic MPO of electrical resistance, also known as phonon-induced resistance
oscillations, have been observed in numerous experiments [1,3,4,6,7,9,13]. They
are well seen under the conditions when phonons play a significant role in relaxation
of electron momentum, for example, in QWs of very high quality [6] where
electron-impurity scattering is minimized, or at elevated temperatures [3]. Measurements
of thermoelectric power (thermopower) in GaAs QWs also show acoustic MPO [2,16], because
the thermoelectric phenomena [17] in GaAs quantum wells are caused mostly by the phonon
drag mechanism [17,18]. The resistance is determined by both electron-phonon
and electron-impurity scattering, the latter prevails at low temperatures. In contrast,
the longitudinal (Seebeck) thermopower due to phonon drag is determined solely by the
electron-phonon scattering, though the electron-impurity scattering remains important for
shaping the density of states of electrons in magnetic field. Therefore, the studies of magnetothermopower is a more direct way for investigation of acoustic MPO as compared to
the studies of magnetoresistance. However, the reported observations [2,16] of these
oscillations in thermopower are very sparse and no detailed comparison of experimental
data to theoretical calculations has been done so far.

In QWs with two or more occupied 2D subbands, there exists another type
of quantum oscillations due to scattering of electrons between the subbands. These
magneto-intersubband oscillations (MISO) [19-34] observed in the resistance measurements
are governed by the {\em magneto-intersubband resonance}, when the difference in subband
energies is a multiple of the cyclotron energy. In these conditions, the impurity-assisted
elastic scattering of electrons between the subbands becomes significant and enhances
the total scattering probability. In two-subband system with subband separation $\Delta$,
the magnetoresistance shows $1/B$-periodic MISO with maxima at $\Delta=n \hbar \omega_c$.
Similar to magnetophonon oscillations, the MISO are robust to increasing temperature,
they were detected at $T$ up to 40 K [24]. The MISO with high amplitudes and large
period are commonly observed in magnetoresistance of double layer structures (such as
double QWs [24] or single wide QWs studied in this paper, see Fig. 1), where subband
separation is small and intersubband scattering is strong.

\begin{figure}[ht]
\includegraphics[width=9.cm]{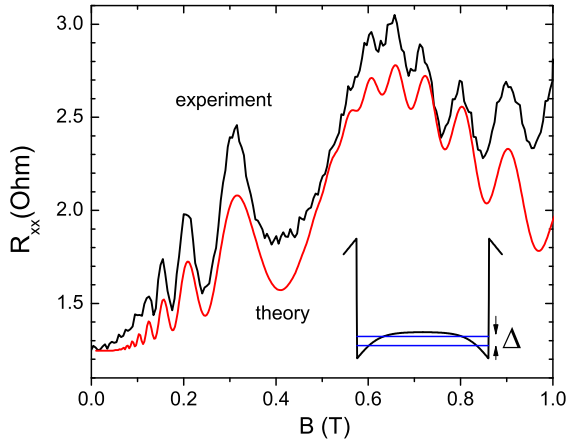}
\caption{(Color online) Magneto-intersubband oscillations of resistance in the wide
QW samples at $T=4.2$ K. A comparison of the measured magnetoresistance with
the calculated one allows us to determine subband separation $\Delta=1.07$ meV
by the oscillation frequency and to estimate the quantum lifetime of electrons
by the amplitude of the oscillations.}
\end{figure}

Recently, measurements of magnetoresistance of a two-subband electron system in a
single QW with high electron density have revealed an interesting phenomenon,
which can be viewed as the interference of acoustic MPO with MISO [9]. In terms
of microscopic quantum processes, such a phenomenon exists because of phonon-assisted intersubband scattering of electrons. Under Landau quantization, this scattering leads to magnetoresistance
oscillations whose periodicity is governed by commensurability of cyclotron energy with
combined energies defined as a sum and a difference of the subband separation energy
$\Delta$ and the characteristic phonon energy $\hbar (k_{1}+k_{2}) s_{\lambda}$, where
$k_{1}$ and $k_{2}$ are the Fermi wavenumbers for subbands 1 and 2. Thus, the {\em combined
resonance} conditions are [9,11]
\begin{eqnarray}
2k_F s_{\lambda} \pm \Delta/\hbar = n \omega_c,
%1
\end{eqnarray}
where $k_F=(k_{1}+k_{2})/2$. The changes in magnetoresistance associated with the
interference in Ref. 9 were definitely resolved near the main magnetophonon resonance
around $B=1$ T. In a wide region of $B$, the behavior of the magnetoresistance was governed
rather by a superposition of MPO and MISO, due to phonon-assisted intrasubband and
impurity-assisted intersubband contributions to transport, respectively.

The interference of MPO with MISO is an important phenomenon because it is a unique
manifestation of interference of two distinct types of quantum magnetooscillations in
quasi-equilibrium macroscopic transport coefficients [35]. However, Ref. 9 still remains
a single report of the observation of the combined resonances (1) in magnetotransport.
In this paper, we propose to employ the measurements of phonon-drag thermopower
as a more convenient method for observation of the combined resonances, compared to the
resistance measurements. The longitudinal thermopower is determined by electron-phonon
scattering and, therefore, does not show up the intersubband resonances due to elastic electron-impurity scattering. This property facilitates detection of the combined resonances
caused by the intersubband phonon-assisted scattering. It is worth noticing that previous
measurements of the thermopower in two-subband electron systems [36-38] were concentrated
on different subjects and did not reveal either the MPO or the combined resonances.

Below we report both experimental and theoretical studies of the magnetothermopower
of a two-subband electron system in a wide (45 nm) GaAs QW. Because of charge redistribution,
a wide QW forms a bilayer (see the inset in Fig. 5), where two wells near the interfaces are
separated by an electrostatic potential barrier, and two subbands appear as a result of tunnel hybridization of 2D electron states. The magnetoresistance of our system shows pronounced
MISO corresponding to $\Delta=1.07$ meV. This value of intersubband separation is close to that
(0.95 meV) obtained from a self-consistent calculation of subband spectrum and wave functions.
While measuring the magnetothermopower, we naturally do not see the MISO, but observe a
considerable change of magnetooscillation picture compared to that in single-subband QWs.
Theoretical calculations satisfactory describe our findings, thereby confirming the importance
of phonon-assisted intersubband scattering in phonon-drag magnetothermoelectric effect.

The paper is organized as follows. Section II describes experimental part and the results.
The details of the theoretical analysis are given in Sec. III. A comparison of the theory with
the experiment, discussion of the results, and concluding remarks are presented in Sec. IV.

\section{Experiment}

We have studied both narrow ($w=14$ nm) and wide ($w=45$ nm) GaAs QWs with electron density
$n_s=6.4 \times 10^{11}$ cm$^{-2}$ and mobility $1.9 \times 10^6$ cm$^2$/V s. The samples
were made in a modified van der Pauw geometry, with an electrically powered heater placed
at the side of the sample, several millimeters away from the 2D layer. The 2D electron gas
occupies a circular central part (diameter 1 mm) and four
long (length 5 mm, width 0.1 mm) arms ending with the voltage probes ( se figure 2). The thermoinduced
voltage $V$ was measured by a lock-in method at the frequency of $2f_0=54$ Hz. The
measurements have been carried out at $T=4.2$ K.
We find the electron temperature near the heater and heat sink by the 2-probe measurements,
exploiting the amplitude of the Shubnikov-de Haas oscillation. The difference in the
electron temperature between hot and cold sides is found $\Delta T \simeq 0.1-0.2$ K at the
lattice temperature $T=4.2$ K. Several devices with narrow and wide QWs from two wafers
have been studied. Figure 2 illustrates magnetic-field dependence of the thermoinduced
voltage for narrow and wide QWs. The voltage increases nearly linearly with heater
power and is almost symmetric with respect to the sign of the magnetic field, which
proves that we measure the longitudinal (Seebeck) thermoelectric effect. In both cases,
we see MPO confirming that the contribution to the thermoelectric effect comes
from the phonon drag mechanism.

\begin{figure}[ht]
\includegraphics[width=9.cm]{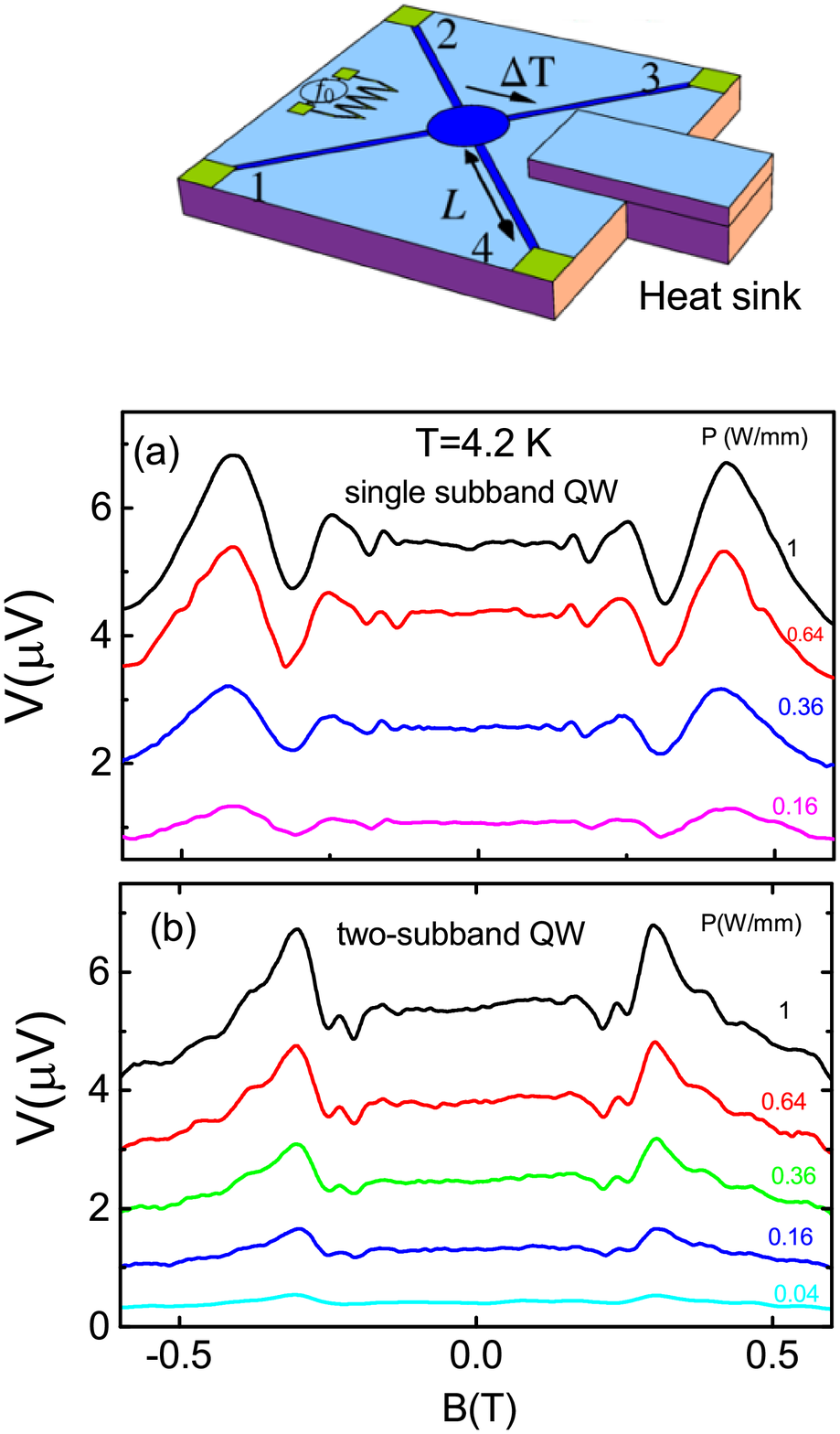}
\caption{(Color online) Sample geometry and magnetic-field dependence of the thermoinduced
voltage for single-subband (a) and two-subband (b) QWs for different heater powers.}
\end{figure}

In Fig.3 we plot the first derivative of the thermopower signal with respect to B, allowing 
to the position of the minima and maxima for 
narrow and wide QWs to be compared. The oscillation position in both systems is coincident only in low magnetic field.
At higher field the position of peak in single QW follows to $1/B$ period, while two-subband system
exhibits several additional oscillations.  
\begin{figure}[ht]
\includegraphics[width=9.cm]{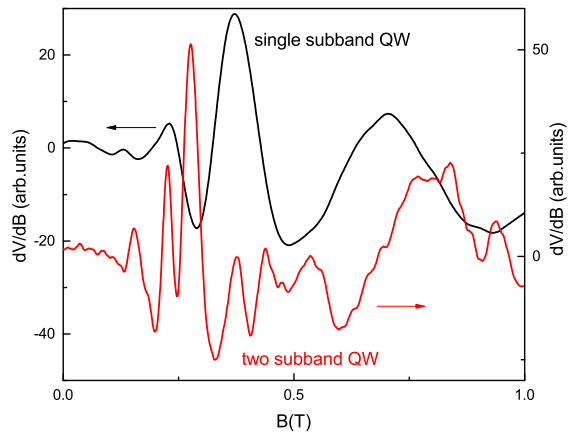}
\caption{(Color online) First derivative of the thermoinduced voltage with respect to B for single and two-subband QWs. }
\end{figure}

Figures 4 and 5 present more detailed plots of the normalized thermoinduced voltage
in single and two-subband QWs for a chosen heater power together with theoretical calculations. In the single-subband
QW the MPO resembles the ones obtained in the previous experiment [2]. In the two-subband
QW we observe a more complicated oscillating picture showing several weaker resonances and
an unexpected growth of the thermoinduced voltage with magnetic field at $B > 0.7$ T. The
theoretical analysis given below allows us to identify these specific for two-subband QWs
features and explain them as a result of intersubband phonon-assisted transitions.

\begin{figure}[ht]
\includegraphics[width=9.cm]{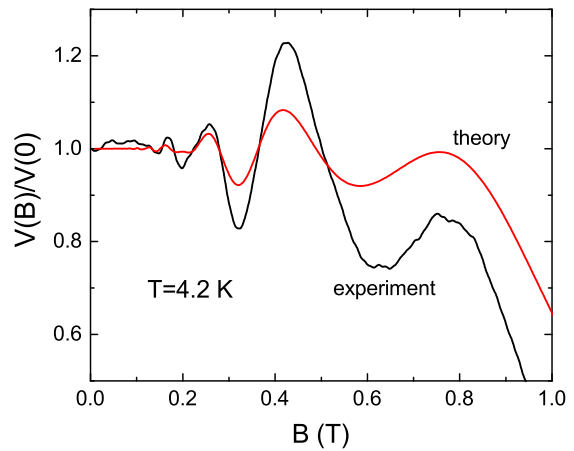}
\caption{(Color online) Magnetic-field dependence of the normalized thermoinduced
voltage for single-subband QW.}
\end{figure}

\begin{figure}[ht]
\includegraphics[width=9.cm]{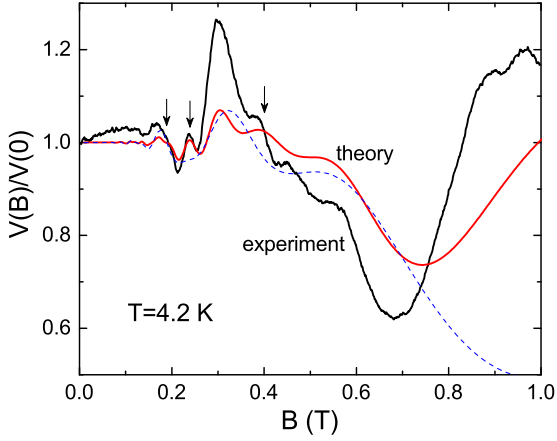}
\caption{(Color online) Magnetic-field dependence of the normalized thermoinduced
voltage for two-subband QW. The combined resonances are marked by the arrows.
The dashed line shows the result of calculation where the intersubband phonon-assisted
scattering was intentionally removed.}
\end{figure}

\section{Theory}

For multi-subband systems, the kinetic theory describing 2D electrons interacting with
impurities and bulk acoustic phonons in the presence of a transverse magnetic field
is developed in Ref. 11. The phonons are described by the mode index
$\lambda$ and wave vector ${\bf Q}=({\bf q},q_z)$, where ${\bf q}$ is the component of
the wave vector in the 2D plane. Under conditions when both cyclotron energy
$\hbar \omega_c$ and phonon energy $\hbar \omega_{\lambda {\bf Q}}$ are much smaller than
the Fermi energy $\varepsilon_F$ (the case of degenerate electron gas, $T \ll \varepsilon_F$,
is assumed), the kinetic equation for the distribution function $f_{j \varepsilon \varphi}$
depending on the subband index $j$, energy $\varepsilon$, and electron momentum angle
$\varphi$ is written as
\begin{eqnarray}
\omega_c \frac{\partial f_{j \varepsilon \varphi}}{ \partial \varphi}  =
J^{im}_{j \varepsilon \varphi}(f) + J^{ph}_{j \varepsilon \varphi}(f),
%2
\end{eqnarray}
where $J^{im}_{j \varepsilon \varphi}$ and $J^{ph}_{j \varepsilon \varphi}$ are the
electron-impurity and electron-phonon collision integrals. The expression for
$J^{ph}_{j \varepsilon \varphi}$ (Eq. (3) of Ref. 11) needs to be generalized to the
case of anisotropic phonon distribution function $N_{\lambda {\bf Q}}$ by substituting
$N_{\lambda {\bf Q}}$ in the phonon emission (first) term and $N_{\lambda -{\bf Q}}$
in the phonon absorption (second) term in place of the isotropic Planck
distribution $N_{\omega_{\lambda {\bf Q}}}$. At temperatures 4.2 K and lower,
the electron-impurity scattering prevails over the
electron-phonon one and controls relaxation of electron momentum in the systems
with mobilities of the order $10^6$ cm$^2$/V s. However, the electron-phonon scattering
is the one responsible for the phonon drag effect. The non-equilibrium part of electron
distribution, $\delta f_{j \varepsilon \varphi}$, appearing due to the drag effect is found
from the equation
\begin{eqnarray}
\omega_c \frac{\partial \delta f_{j \varepsilon \varphi}}{ \partial \varphi}  =
\delta J^{ph}_{j \varepsilon \varphi}(f^{(0)})+ J^{im}_{j \varepsilon \varphi}(\delta f),
%3
\end{eqnarray}
where $f^{(0)}_{j \varepsilon}$ is the equilibrium Fermi distribution function and
$\delta J^{ph}_{j \varepsilon \varphi}$ is the contribution to collision integral
caused by the antisymmetric in ${\bf Q}$ part of the phonon distribution
function, $\delta N_{\lambda {\bf Q}}$:
\begin{eqnarray}
\delta J^{ph}_{j \varepsilon \varphi}= \frac{m}{\hbar^3}
\sum_{j'} \int_0^{2 \pi} \frac{d \varphi'}{2 \pi}
\sum_{\lambda} \int_{-\infty}^{\infty} \frac{d q_z}{2\pi} C_{\lambda {\bf Q}_{jj'}} I_{jj'}(q_z)
\nonumber \\
\times \delta N_{\lambda {\bf Q}_{jj'}} \sum_{l=\pm 1} l D_{j' \varepsilon - l \hbar \omega_{\lambda {\bf Q}_{jj'} }}
(f^{(0)}_{j' \varepsilon - l \hbar \omega_{\lambda {\bf Q}_{jj'} }} -f^{(0)}_{j \varepsilon}).
%4
\end{eqnarray}
The phonon wave vector ${\bf Q}_{jj'}=({\bf q}_{jj'},q_z)$ in this expression depends on the
subband indices. Its in-plane component ${\bf q}_{jj'}$ is defined by the polar angle
$\varphi_q=\arctan [ (k_j\sin \varphi-k_{j'} \sin \varphi')/(k_j\cos \varphi-k_{j'}\cos \varphi')]$
and absolute value $q_{jj'}= \sqrt{k^2_{j} + k^2_{j'} - 2 k_{j}k_{j'} \cos \theta}$, where
$k_{j}$ is the Fermi wavenumber in the subband $j$ and $\theta=\varphi-\varphi'$ is the
scattering angle. Owing to smallness of phonon energies, the quasielastic scattering
approximation used in these expressions is justified. Next, $D_{j \varepsilon}$ is
the density of states in subband $j$, expressed in units of $m/\pi \hbar^2$,
$C_{\lambda {\bf Q}_{jj'}}$ is the squared matrix element of electron-phonon
interaction in the bulk, determined by both deformation-potential and piezoelectric
mechanisms of the interaction, and
\begin{eqnarray}
I_{jj'}(q_z)=\left|\int dz \Psi^*_j(z) e^{iq_z z} \Psi_{j'}(z) \right|^2
%5
\end{eqnarray}
is the overlap factor determined by the envelope wave functions $\Psi_{j}(z)$ and
$\Psi_{j'}(z)$ of the corresponding subbands. For a wide quantum well, these factors
are to be calculated numerically (Fig. 5).

\begin{figure}[ht]
\includegraphics[width=9.cm]{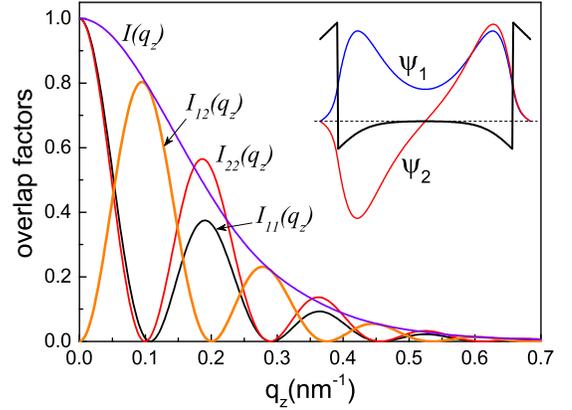}
\caption{(Color online) Overlap factors $I_{jj'}$ and their combination
$I=(I_{11}+I_{22})/2 +I_{12}$ for QW of width $w=45$ nm with density $n_s=6.4 \times
10^{11}$ cm$^{-2}$ studied in our experiment. The inset shows the confinement potential
and wave functions of the first (symmetric) and the second (antisymmetric) subband states
determined by a self-consistent solution of the Schroedinger and Poisson equations.}
\end{figure}

Since $J^{im}_{j \varepsilon \varphi}(\delta f)$ is linear in $\delta f_{j \varepsilon \varphi}$,
the kinetic equation (3) with $\delta J^{ph}_{j \varepsilon \varphi}$ of Eq. (4) is solved
straightforwardly. The thermoelectric current density,
\begin{eqnarray}
{\bf j}_T =\frac{e}{\pi \hbar} \sum_j \int d \varepsilon D_{j \varepsilon} k_j \int_0^{2 \pi}
\frac{d \varphi}{2 \pi}
\left( \begin{array}{c} \cos \varphi \\ \sin \varphi \end{array} \right)
\delta f_{j \varepsilon \varphi},
%6
\end{eqnarray}
is determined by $\delta f_{j \varepsilon \varphi}$. Because of high mobility
of electrons, we consider the regime of classically strong magnetic fields,
when $\omega_c$ is much larger than the inverse transport scattering time $1/\tau_{tr}$.
The zero-order iteration in small parameter $1/\omega_c \tau_{tr}$ is sufficient
for calculation of the longitudinal thermopower. The thermoelectric current in
this approximation is independent of scattering by impurities:
\begin{eqnarray}
{\bf j}_T \simeq \frac{e}{m \omega_c} \sum_{jj'} (n_j+n_{j'})  \nonumber \\
\times \hat{\cal G}_{jj'} \biggl\{  q_{jj'}^{-1} \delta N_{\lambda {\bf Q}_{jj'}} R_{jj'}
\left( \begin{array}{c} -\sin \varphi_q \\ \cos \varphi_q \end{array} \right) \biggr\},
%7
\end{eqnarray}
where $n_j=k^2_j/2 \pi$ is the electron density in the subband $j$,
$\widehat{\cal G}_{jj'}$ is the integral operator defined as
\begin{eqnarray}
\widehat{\cal G}_{jj'} \{ A \} \equiv \frac{2m^2}{\hbar^3} \int_0^{2 \pi} \frac{d \theta}{2 \pi}
\int_0^{2 \pi} \frac{d \varphi_q}{2 \pi} \sum_{\lambda} \int_{-\infty}^{\infty} \frac{d q_z}{2\pi} \nonumber \\
\times \omega_{\lambda {\bf Q}_{jj'}} C_{\lambda {\bf Q}_{jj'}} I_{jj'}(q_z) {\cal F}_{jj'}(\theta) A,
%8
\end{eqnarray}
${\cal F}_{jj'}(\theta)=1 - 2 k_j k_{j'} \cos \theta/(k_j^2 + k_{j'}^2)$, and
\begin{eqnarray}
R_{jj'} = \frac{1}{2 \hbar \omega_{\lambda {\bf Q}_{jj'}}}\int d \varepsilon
(f^{(0)}_{\varepsilon-\hbar \omega_{\lambda {\bf Q}_{jj'}} } - f^{(0)}_{\varepsilon} ) \nonumber \\
\times \left( D_{j \varepsilon}
D_{j' \varepsilon-\hbar \omega_{\lambda {\bf Q}_{jj'}} }  + D_{j' \varepsilon}
D_{j \varepsilon-\hbar \omega_{\lambda {\bf Q}_{jj'}} } \right).
%9
\end{eqnarray}

Finally, specifying the phonon distribution as [39]
\begin{eqnarray}
\delta N_{\lambda {\bf Q} }= \frac{\partial N_{\omega_{\lambda {\bf Q}}}}{\partial
\omega_{\lambda {\bf Q}}} \frac{\omega_{\lambda {\bf Q}}}{T} \tau_{\lambda} {\bf u}_{
\lambda {\bf Q}} \cdot \nabla T,
%10
\end{eqnarray}
where $\tau_{\lambda}$ is the phonon lifetime and ${\bf u}_{\lambda {\bf Q}}=\partial
\omega_{\lambda {\bf Q}}/\partial {\bf Q}$ is the phonon group velocity,
one can find the current density in the standard form ${\bf j}_T=-\hat{\beta} \nabla T$,
where the thermoelectric tensor $\hat{\beta}$, in view of the assumed condition
$\omega_c \tau_{tr} \gg 1$, has only non-diagonal components. The longitudinal
thermopower is given by the following expression
\begin{eqnarray}
\alpha_{xx} \simeq \rho_{xy} \beta_{yx} = -\frac{1}{\hbar |e|} \sum_{jj'}
\frac{n_j+n_{j'}}{2 n_s} \nonumber \\
\times \hat{\cal G}_{jj'} \biggl\{ \tau_{\lambda}  Q_{jj'}^{-2}
F \left(\frac{\hbar \omega_{\lambda {\bf Q}_{jj'}} }{2T} \right) R_{jj'} \biggr\},
%11
\end{eqnarray}
where $F(x)=[x/\sinh(x)]^2$. The Hall resistance is $\rho_{xy}=m \omega_c/e^2 n_s$. In
the case of single subband occupation, the expression for $\beta_{yx}$ is reduced to
the one obtained in Ref. 15. Calculation of the integral over energy in Eq. (9)
is considerably simplified under condition $2 \pi^2 T \gg \hbar \omega$, when Shubnikov-de
Haas oscillation are thermally suppressed. At $T=4.2$ K this condition is satisfied up to
$B=1$ T, so we use it in the following.

Below, for analysis of experimental data, we restrict ourselves by the approximation
of overlapping Landau levels, when only the first oscillatory harmonics of the density
of states are taken into account: $D_{j \varepsilon} \simeq 1 - 2 d_j \cos[2 \pi
(\varepsilon-\varepsilon_j)/\hbar \omega_c]$, where $d_j$
are the Dingle factors and $\varepsilon_j$ are the quantization energies of the subbands.
Since the subband separation $\Delta=\varepsilon_2-\varepsilon_1$ is much smaller than
$2\varepsilon_F$, we also neglect the difference between $k_1$ and $k_2$. The latter
approximation means that $q_{jj'} \simeq 2 k_F \sin(\theta/2)$ (so that ${\bf Q}_{jj'}$ is
no longer dependent on the subband indices), $n_1 \simeq n_2 \simeq n_s/2$ and
${\cal F} \simeq 1-\cos \theta$. Thus, calculation of the sum over the subband
indices in Eq. (11) is reduced to calculation of the factor
\begin{eqnarray}
{\cal I}_{\lambda {\bf Q}}(q_z) = \frac{1}{2} \sum_{jj'} I_{jj'}(q_z) R_{jj'} \simeq
I(q_z) + \cos \frac{2 \pi \omega_{\lambda {\bf Q}} }{\omega_c} \nonumber \\
\times \left(d_1^2 I_{11}(q_z)+d_2^2 I_{22}(q_z) + 2 d_1d_2 I_{12}(q_z)
\cos \frac{2 \pi \Delta}{\hbar \omega_c} \right),
%12
\end{eqnarray}
where $I(q_z)=(I_{11}+I_{22})/2 +I_{12}$. The thermopower takes the form
\begin{eqnarray}
\alpha_{xx} \simeq -\frac{m^2}{|e|\hbar^4} \int_0^{2 \pi} \frac{d \theta}{2 \pi}
\int_0^{2 \pi} \frac{d \varphi_q}{2 \pi}
\sum_{\lambda} \int_{0}^{\infty} \frac{d q_z}{\pi} \nonumber \\
\times (1-\cos \theta) C_{\lambda {\bf Q}}
\tau_{\lambda} F\left(\frac{\omega_{\lambda {\bf Q}}}{2T} \right)
\frac{2 \omega_{\lambda {\bf Q}}}{Q^2} {\cal I}_{\lambda {\bf Q}}(q_z).
%13
\end{eqnarray}
Comparing Eq. (13) with the results of Ref. 15, one may notice that
the single-subband case is described by the substitution ${\cal I}_{\lambda
{\bf Q}} \rightarrow I_{q_z}[1 + 2 d^2 \cos(2 \pi \omega_{\lambda {\bf Q}}/\omega_c)]$,
where $I_{q_z}$ is the corresponding overlap factor.

The expression for ${\cal I}_{\lambda {\bf Q}}(q_z)$ comprises both the classical
contribution proportional to $I(q_z)$ and the quantum contribution containing MPO
originating from intrasubband (terms at $I_{11}$ and $I_{22}$) and intersubband (term
at $I_{12}$) transitions of electrons. The presence of the product of magnetophonon
oscillating factor $\cos(2 \pi \omega_{\lambda {\bf Q}}/\omega_c)$ by the
magneto-intersubband oscillating factor $\cos(2 \pi \Delta/\hbar \omega_c)$ formally
discloses the interference nature of the intersubband term. This product is also
representable as a sum of oscillating factors with the combined frequencies
$\omega^{\pm}_{\lambda {\bf Q}}=\omega_{\lambda {\bf Q}} \pm \Delta/\hbar$ leading to
combined resonances according to Eq. (1).

\section{Numerical results and discussion}

The results of numerical calculation of the normalized thermopower
$\alpha_{xx}(B)/\alpha_{xx}(0)=V(B)/V(0)$
for both single-subband and two-subband QWs are presented together with the experimental plots in
Figs. 3 and 4. The calculation for two-subband QW are done according to Eqs. (12) and (13),
under approximation $d_1=d_2=\exp(-\pi/ \omega_c \tau)$, where $\tau$ is the quantum lifetime
of electrons, common for both subbands.

%The calculation for single-subband QW has been done using the density of states
%determined in the self-consistent Born approximation. The result shows only a
%slight difference (at large $B$) from the one based upon the approximation of
%overlapping Landau levels, thereby justifying the reliability of this approximation
%in the region $B < 1$ T.

In the calculations, we used the parameters of our samples together with material
parameters of GaAs, substituted $\tau=7$ ps (estimated from the MISO amplitude, see
Fig. 1), and assumed $\tau_{\lambda}$ as a mode-independent constant. We need to
emphasize that the theory is based on the model form of non-equilibrium part of the
phonon distribution function, Eq. (10), while the actual phonon distribution in our
experiment is influenced by geometrical details [16] and, therefore, may considerably
deviate from this form. For this reason, we do not expect a good
agreement between the theory and the experiment as concerns the amplitudes of the
oscillations. On the other hand, both the general behavior of the thermoinduced
voltage and the position of extrema should be reproduced correctly, and we indeed
have this kind of agreement. However, we noticed that the contribution of the
transverse phonon modes to the thermoelectric effect is larger than expected from
the theory. Since the interaction with transverse modes is associated mostly with
the piezoelectric mechanism of electron-phonon interaction, we used the piezoelectric
potential constant $h_{14}$ as a single adjustable parameter, to find a better
correspondence between the theory and the experiment. We have found that both
single-subband and two-subband cases are described considerably better when $h_{14}$
is increased from the usually assumed value of $1.2$ V/nm to $2.8$ V/nm. The increased
importance of transverse modes and/or piezoelectric mechanism in the thermoelectric
experiments can be likely associated with excitation of surface acoustic phonon
modes as a result of external heating. These phonons interact with 2D electrons mostly
via the long-range piezoelectric fields and their frequency is close to the frequencies
of the bulk transverse phonons [40]. The problem of the possible contribution of surface
acoustic phonons to magnetothermopower oscillations requires a special study which is
beyond the scope of the present work.

The characteristic features appearing in the thermoelectric effect in two-subband
QWs because of intersubband phonon-assisted scattering deserve a discussion.
Since the intersubband separation $\Delta$ in our samples is comparable to resonance
phonon frequencies $\omega_{ph} = 2k_F s_{\lambda}$, the most prominent manifestation
of the combined resonances is the appearance of extra peaks when the MPO minima
coincide with MISO minima. In other words, when both $\omega_{ph}$ and $\Delta/\hbar$
are half-integer multiples of $\omega_c$, the combined frequencies $\omega_{ph} \pm
\Delta/\hbar$ are integer multiples of $\omega_c$, so instead of a local minimum
expected in the absence of intersubband transitions one has a local maximum. We
clearly observe this kind of peaks at $B \simeq 0.24$ T and $B \simeq 0.4$ T. There
are also weaker features such as a barely resolved local minimum at $B \simeq 0.18$ T.
When the MPO maxima coincide with MISO maxima (both $\omega_{ph}$ and $\Delta/\hbar$
are integer multiples of $\omega_c$), the enhancement of the thermoinduced voltage
takes place, for example, at $B \simeq 0.3$ T.

The non-monotonic behavior of the thermoinduced voltage at higher fields is also a
consequence of intersubband scattering (compare two theoretical plots in Fig. 4).
When $\omega_c$ becomes larger than the resonance frequency of the highest-energy
phonon mode (in our sample, when $B > 0.55$ T), the probability of phonon-assisted
scattering within the same subband decreases monotonically. However, because of
enhancement of the phonon-assisted scattering between the adjacent Landau levels of
different subbands (these levels are separated by the energy $\hbar \omega_c - \Delta$),
the thermoinduced voltage passes through a minimum near $B=0.7$ T and then increases.

The intersubband scattering of electrons requires phonons with finite perpendicular
component of the wave vector, $q_z$. On the other hand, the magnetophonon resonance
occurs at $q_z$ much smaller than $2 k_F$. In wide QWs, where $k_F w \gg 1$, these
conditions do not contradict with each other, because the overlap factor $I_{12}(q_z)$
becomes sufficiently large already at $q_z \ll 2 k_F$. The presence of non-equilibrium
phonons with finite $q_z$ can be explained even in the case of ballistic phonon
propagation: such phonons coming from the heater reflect from the upper and lower
boundaries of the sample and can reach the 2D layer.

In conclusion, we observe acoustic magnetophonon oscillations of thermopower in
high-mobility 2D electron gas in quantum wells with two occupied 2D subbands and
detect the combined resonances caused by intersubband phonon-assisted transitions
of electrons. A detailed comparison of experimental and theoretical dependence of
thermopower on magnetic field is carried out. The importance of intersubband
transitions in the phonon drag effect in magnetic field is demonstrated.\\

The financial support of this work by FAPESP, CNPq (Brazilian agencies) is acknowledged.


\begin{thebibliography}{40}
\bibitem{35}
Other quantum interference phenomena are observed in nonlinear magnetotransport
at strong dc driving and in non-equilibrium magnetotransport under microwave
irradiation (see Ref. 14 for a review).
\end{thebibliography}
\end{document}